\documentclass[12pt]{l4dc2020} 
\newtheorem{assumption}{Assumption}

\usepackage{graphicx}
\usepackage{wrapfig}

\usepackage[compact]{titlesec}         

\usepackage{soul}
\usepackage{xcolor}
\usepackage{appendix}
\usepackage{cleveref}

\crefname{equation}{}{} 

\usepackage{mathtools}
\usepackage{algorithm}
\usepackage{algpseudocode}

\def\lone{{\mathcal{L}_1}}
\def\lonew{${\mathcal{L}_1}$ }
\def\linf{{\mathcal{L}_\infty}}
\def\linfw{$\mathcal{L}_\infty$ }

\def\mI{\mathbb{I}}

\def\mX{\mathbb{X}}

\newcommand{\norm}[1]{\left\lVert#1\right\rVert}

\newcommand{\pr}{\text{Pr}}

\title[$\mathcal{L}_1$ Adaptive Control with GPR]{$\mathcal{L}_1$-$\mathcal{GP}$: $\mathcal{L}_1$ Adaptive Control with Bayesian Learning 
}    

\usepackage{times}

\author{%
 \Name{Aditya Gahlawat} \thanks{Mechanical Science and Engineering, University of Illinois at Urbana Champaign, Urbana, IL-61801} \Email{gahlawat@illinois.edu}
 \vspace{-0.2cm}
  \AND
 \Name{Pan Zhao}\footnotemark[1] \Email{panzhao2@illinois.edu}
  \vspace{-0.2cm}
 \AND
 \Name{Andrew Patterson}\footnotemark[1] \Email{appatte2@illinois.edu}
  \vspace{-0.2cm}
 \AND
 \Name{Naira Hovakimyan}\footnotemark[1] \Email{nhovakim@illinois.edu}
  \vspace{-0.2cm}
 \AND
 \Name{Evangelos A. Theodorou} \thanks{Aerospace Engineering, Georgia Institute of Technology, Atlanta, GA-30332} \Email{evangelos.theodorou@ae.gatech.edu}
}

\begin{document}
\maketitle

\vspace{-10mm}

\begin{abstract}%
We present $\mathcal{L}_1$-$\mathcal{GP}$, an architecture based on $\mathcal{L}_1$ adaptive control and Gaussian Process Regression (GPR) for {\it safe simultaneous control and learning}. 
On  one hand, the \lonew adaptive control provides stability and transient  performance guarantees, which allows for GPR to efficiently and safely learn the uncertain dynamics. On the other hand, the learned dynamics can be conveniently incorporated into the \lonew control architecture without sacrificing robustness and tracking performance. Subsequently, the learned dynamics can lead to less conservative designs for performance/robustness tradeoff. We illustrate the efficacy of the proposed architecture via numerical simulations.     

\end{abstract}



\section{Introduction}
The historical premise of adaptive control was to control uncertain systems while simultaneously learning the system parameters and providing robustness to uncertainties. Rudolf Kalman was the first to coin the term ``self-tuning controller'' in 1958 by introducing optimal linear-quadratic regulator (LQR) with explicit identification of parameters \citep{kalman1958self}. The field of adaptive control since then witnessed tremendous developments, capturing different classes of nonlinear systems, including presence of unmodeled dynamics, switching models, hybrid systems and other singularities, e.g. \citet{astrom2008adaptive,landau79adaptive,narendra1980stable,sastry2011adaptive,ioannou2012robust}, and references therein. The main architectures were inspired by inverse Lyapunov design, ensuring asymptotic stability in the presence of system uncertainties and disturbances. Recent developments in $\mathcal L_1$ adaptive control filled the last gap of explicitly introducing robustness into the problem formulation, leading to a framework with \textit{a priori} guaranteed robustness margins,   transient and steady-state specifications \citep{cao2008TAC,L1_book}. In \lonew control architecture, estimation is decoupled from control, thereby allowing for arbitrarily fast adaptation subject  only to hardware limitations. The \lonew control has been successfully implemented on NASA's AirStar 5.5\% subscale generic transport aircraft model \citep{gregory2009l1, gregory2010flight} and~Calspan's Learjet \citep{GNC16_L1Learjet,ackerman2017evaluation} and F16 aircraft 
and unmmaned aerial vehicles  \citep{kaminer2010path,kaminer2015time,jafarnejadsani2017optimized,zuo2014augmented}. 
Despite these vast developments, the issue of learning the system dynamics and/or uncertainties remained unresolved, as the typical estimation schemes in all these adaptive architectures require persistency of excitation (PE) type assumption on reference signals to ensure parameter convergence. Such requirement is unacceptable in safety-critical applications, rendering the conventional Lyapunov-based adaptive control architectures incomplete, if parameter/system identification is to be addressed {\it simultaneously} with  transient specifications.



The last two decades have witnessed a type of data explosion that has revolutionized the industry of autonomous systems. Tools from machine learning have been extensively explored in modeling, identification, and control of dynamic systems. A few examples of such tools include, but are not limited to, neural networks~\citep{lewis1998neural}, Gaussian processes~\citep{williams2006gaussian}, and reinforcement learning \citep{sutton2018reinforcement}. In many of these instances, guarantees of stability have not been prioritized, yet having an impressive demonstration was the main objective to show the power of data-driven methods towards achieving full autonomy   \citep{lillicrap2015continuoousRL,deisenroth2013gaussian,levine2016end,pan2019imitation}. Due to its data efficiency, the nonparameteric structure and the ability to provide uncertainty quantification, Gaussian Process Regression (GPR) has become popular in safety-critical learning and control \citep{aswani2013provably,akametalu2014reachability,berkenkamp2015safe,berkenkamp2017safe,hewing2019cautious,wang2018safe}, including application to model reference adaptive control \citep{chowdhary2014bayesian}. When the learning methods generate unsafe reference trajectories, the control barrier function methods presented by~\citet{cheng19end2end} and \citet{salehi2019active} correct the control input to ensure the system state remains in a safe set. This approach assumes that the reference trajectory may be unsafe or infeasible. In the present work the desred trajectory is designed to be feasible and safe for an appropriately designed reference system. The safety and feasibility guarantees are then dependent on the ability of an adaptive-controller to emulate the reference system. This design philosophy allows safe and feasible trajectories to be generated \textit{a priori}, instead of relying on run-time optimization routines to correct the unsafe trajectories.

However, in most of the techniques presented, the control performance is a direct function of the  quality of the learned uncertainties. The method presented by~\citet{taylor2019adaptive} specifically considers performance and uses an adaptive controller to ensure asymptotic tracking performance while avoiding unstable reference commands. In this paper we combine the formal stability and robustness guarantees of $\mathcal{L}_1$ adaptive control with Gaussian Processes to ensure safe learning and adaptation with \textit{a priori} transient bounds.  
This would enable the satisfaction of control objectives like trajectory tracking and simultaneously enable learning from the collected data.




Over the last two years \lonew control has been explored within NASA’s Learn-To-Fly (L2F) framework. In this work, a real-time system identification toolbox of NASA is integrated across the flight envelope to continuously update the model parameters and enable autonomous flight without intensive wind-tunnel testing, while an \lonew adaptive controller is used to provide robustness and stability guarantees~\citep{snyder2019phd}.
Incorporation of learning via neural network in \lonew control was investigated in \citet{cooper2014learning}.
The system identification within L2F and/or the neural network based learning require some prior knowledge of the system and/or uncertainty structure to facilitate parameter estimation. 

In this paper we explore the \lonew control architecture with Bayesian learning in the form of GPR for safe learning with guaranteed stability and control performance throughout the learning phase.
We assume no availability of model structure and resort to the GPR to learn the uncertain dynamics  whenever possible, while achieving given control objectives like trajectory tracking. 
The predictor in \lonew adaptive control architecture naturally allows the incorporation of the availabe knowledge in a systematic way\footnote{The \textit{apriori} knowledge of a system such as time-delay and input saturation can be conveniently incorporated into the state predictor, which helps to improve both the performance and robustness \citep{Evgeny2011-L1Limiting}.}. 
We demonstrate that one can learn model uncertainties efficiently and safely via GPR, while guaranteeing the stability and performance. Furthermore, we  illustrate that the fast adaptation of \lonew controller intervenes when the uncertainties change. This ensures safe control while the Bayesian learning catches up. 

Finally, one may argue that if $\mathcal{L}_1$ adaptive control already guarantees stability and robustness, then why incorporate learning within it. Instead, the learning should be kept separate if the goal is just safe learning.  While this assertion is true, in addition to safe learning, we are also demonstrating that learning can be incorporated within the $\mathcal{L}_1$ architecture without harming robustness or performance. This is the initial step of the envisioned research, where the next step is to illustrate how the learning can  improve   performance, without sacrificing robustness, when a larger operational envelope is considered as compared to a single trim condition. On the other hand, the benefits of $\mathcal{L}_1$-$\mathcal{GP}$ for purposes of planning (guidance and navigation) in highly uncertain environments are yet to be illustrated on appropriate benchmark examples. 

 The paper is organized as follows. The problem formulation is introduced in Section~\ref{sec:problem_statement}, and and overview of Bayesian learning via GPR and \lonew adaptive control is  provided. The main architecture of $\lone-\mathcal{GP}$ is presented in Section~\ref{sec:L1-GP}. Numerical validation of the proposed architecture is demonstrated in Section~\ref{sec:sims}. The manuscript is concluded in Section~\ref{sec:conclsn}.



\section{Problem Formulation}
\label{sec:problem_statement}
We start this section by providing the notation used in our paper. In particular, let $\|\cdot\|_p$ denote the $p$-norm defined on the space $\mathbb{R}^n$ and  $n \in \mathbb{N}$, and  $\|\cdot\|$ denote the $2$-norm. $\mI_n$ denotes an identity matrix of size $n$. Given a positive scalar $\kappa$, we denote by $\mathbb{X}_\kappa$ the compact set containing all $x \in \mathbb{R}^n$ such that $\|x\|_\infty \leq \kappa$.
Similarly, arbitrary compact subsets of $\mathbb{R}^n$ are denoted by $\mathbb{X}$. For any time-varying function $g(t)$, $g(s)$ denotes its Laplace transform when it exists, and $\norm{g}_\linf$ denotes its \linfw  norm. For a transfer function matrix $G(s)$, $\norm{G(s)}_{\lone}$ denotes its \lonew-norm.    Next we discuss the problem formulation by  considering the following system:  
\begin{equation}\label{eqn:system_dynamics}
\dot{x}(t) = A_m x(t) + B_m(u(t)+f(x(t))), \quad x(0) = x_0, \quad \text{and} \quad y(t) =  C_m x(t),
\end{equation}
 where $x(t) \in \mathbb{R}^n$ is the system state,  $u(t) \in \mathbb{R}^m$ is the control input, $A_m \in \mathbb{R}^{n \times n}$ is a known Hurwitz matrix  specifying the desired closed-loop dynamics, $B_m \in \mathbb{R}^{n \times m}$ and $C_m \in \mathbb{R}^{m \times n}$, $m \leq n$,
 are known matrices with $\text{rank}(B_m) = m$, $f:\mathbb{R}^n \rightarrow \mathbb{R}^m$ is the \emph{unknown} nonlinearity representing the model uncertainties, and $y(t) \in \mathbb{R}^m$ is the regulated output. The matrices $A_m$, $B_m$ and $C_m$ are the designed reference system matrices and express the desired closed-loop system behavior.
\vspace{-2mm}
\begin{assumption}\label{assmp:kernel}
The constituent functions of the unknown nonlinearity $f = \begin{bmatrix} f_i & \cdots & f_m  \end{bmatrix}^\top$, $f_i:\mathbb{R}^n \rightarrow \mathbb{R}$ are samples from  Gaussian processes $\mathcal{GP}(0,K_{f,i}(x,x'))$, where the kernels $K_{f,i}:\mathbb{R}^n \times \mathbb{R}^n \rightarrow \mathbb{R}$ are known. Furthermore, we assume that the kernels are Lipschitz on compact subsets of $\mathbb{R}^n$ with known Lipschitz constants $L_{k,i}(\mathbb{X})$. 
\end{assumption}
\vspace{-6mm}
\begin{assumption}\label{assmp:Lip_bounds}
There exists a known conservative bound $L_f(\mathbb{X})$ such that $\left\| \nabla_x f(x)    \right\|_\infty \leq  L_f(\mathbb{X})$ for all  $x \in \mathbb{X}$,
and $B_0$ such that 
 $\norm{f(0)}_\infty \leq B_0$.

\end{assumption}

The objective is to  learn the model uncertainty $f$ and  track a given bounded reference signal $r(t)$ with quantifiable performance bounds both in transient and steady-state. Next we discuss the two ingredients of our approach, namely GPR and $\mathcal{L}_1$ adaptive control.





\subsection{Bayesian Learning of  Model Uncertainties}

We present the high-probability bounds for the uniform prediction errors by first setting up the measurement model. Assume we have $N \in \mathbb{N}$ measurements of the form
\[
y_j =
 f(x_j) + \zeta =
\left(B_m^\top B_m  \right)^{-1} B_m^\top \left(\dot{x}_j - A_m x_j  \right) - u_j + \zeta \mathbb{I}_m, \quad \zeta  \sim \mathcal{N}(0,\sigma_n^2), \quad y_j \in \mathbb{R}^m,
\] where $j \in \{1,\dots,N\}$ and $\zeta$ is a zero-mean i.i.d. Gaussian random variable representing measurement noise. 
Note that we usually only have access to measurements of $x$ and $u$, and not $\dot{x}$. However, estimates of $\dot{x}$ may be numerically generated with the estimation errors incorporated into $\zeta$. As an example, one may use the Savitsky-Golay filter for this purpose,~\citep{schafer2011savitzky}. 
Using the measurements, we define the data set as $\mathcal{D}_N  = \{\mathbf{Y},\mathbf{X}\}$,
where $\mathbf{Y} \in \mathbb{R}^{N \times m}$, $\mathbf{X} \in \mathbb{R}^{N \times n}$ and are defined as $\mathbf{Y} = \begin{bmatrix} y_1 & \cdots & y_N \end{bmatrix}^\top$, and $\mathbf{X} = \begin{bmatrix} x_1 & \cdots & x_N \end{bmatrix}^\top$. Note that the boldface matrices are directly dependent on the observed data.
GPR proceeds by using the assumption that $f_i \sim \mathcal{N}(0,K_{f_i}(x,x'))$, $i \in \{1,\dots,m\}$, and the data $y_j \sim \mathcal{N}(f(x_j),\sigma_n^2 \mathbb{I}_m)$ to formulate the posterior distributions conditioned on data at any test point $x^\star \in \mathbb{R}^n$ as
\begin{equation}\label{eqn:posterior_conditionals}
    f_i(x^\star)|\mathbf{Y}_i \sim \mathcal{N}(\mu_i(x^\star),\sigma_i^{2}(x^\star)), \quad i \in \{1,\dots,m\},
\end{equation} where $\mathbf{Y}_i$ is the $i^{\text{th}}$ column of $\mathbf{Y}$.  The terms  $ \mu_i(x^\star)  $  and $ \sigma_i(x^\star) $ are mean and variance of the GP model 
and are defined as  $\mu_i(x^\star) = \mathbf{K}_i^{\mathbf{\star}}(x^\star)^\top  \left( \mathbf{K}_i + \sigma_n^2 \mathbb{I}_N \right)^{-1}\mathbf{Y}_i $,  
  and
   $ \sigma_i^{2}(x^\star) = \mathbf{K}^\mathbf{\star \star}_i(x^\star) - \mathbf{K}_i^{\mathbf{\star}}(x^\star)^\top  \left( \mathbf{K}_i + \sigma_n^2 \mathbb{I}_N \right)^{-1}\mathbf{K}_i^{\mathbf{\star}}(x^\star)$. 
  The terms  $ \mathbf{K}_i^{\mathbf{\star\star}}(x^\star) $,  $\mathbf{K}_i^{\mathbf{\star}}(x^\star) $ and $\mathbf{K}_i $ are defined based on the kernel of GP model as $
    \mathbf{K}_i^{\mathbf{\star\star}}(x^\star) = K_{f,i}(x^\star,x^\star) \in \mathbb{R}, \quad
\mathbf{K}_i^{\mathbf{\star}}(x^\star) = K_{f,i}(\mathbf{X},x^\star) \in \mathbb{R}^N, \quad \mathbf{K}_i = K_{f,i}(\mathbf{X},\mathbf{X}) \in \mathbb{R}^{N \times N}$.
Further details can be found in~\citet{williams2006gaussian} and~\citet{bishop2006pattern}. 
A major advantage of GPR is that the predictive estimates are in the form of predictive distributions, as in~\eqref{eqn:posterior_conditionals}, as opposed to point estimates. These predictive distributions can be used to produce high probability bounds on the prediction errors. For example,~\citet{srinivas2012information,chowdhury2017kernelized} present methods of computing uniform prediction error bounds in the context of GP-optimization. These bounds are information-theoretic, which make them generally difficult to compute, especially in an on-line setting. Recently, the authors in~\citet{lederer2019uniform} presented a method of computing similar bounds, which are amenable to on-line computation. The following result is a generalization of~\citet[Thm.~3.1]{lederer2019uniform}.
\vspace{-2mm}

\begin{theorem}\label{theorem:uniform_bounds} 
Let the model uncertainty $f$ satisfy Assmuptions~\ref{assmp:kernel}-~\ref{assmp:Lip_bounds}. Given the posterior distributions in~\eqref{eqn:posterior_conditionals},  for some $\xi > 0$ and any compact set $\mathbb{X} \subset \mathbb{R}^n$, let 
\begin{align*}
\mu(x) =& \begin{bmatrix}
\mu_1(x) & \cdots & \mu_m(x)
\end{bmatrix} , \quad \sigma(x) = \begin{bmatrix}
\sigma_1(x) & \cdots & \sigma_m(x) 
\end{bmatrix}, \\
L_{\mu_i}(\mathbb{X}) =& L_{k,i}(\mathbb{X})\sqrt{N}\|(\mathbf{K}_i + \sigma_n^2 \mathbb{I}_N)^{-1} \mathbf{Y}_i\|,\\
\omega_{\sigma_i}(\xi) = & \sqrt{2 \xi L_{k,i}(\mathbb{X}) \left(1 +  N  \|(\mathbf{K}_i + \sigma_n^2 \mathbb{I}_N)^{-1}\| \max_{x,x' \in \mathbb{X}}K_{f,i}(x,x')\right)},\\
L_\mu(\mathbb{X}) = & \max_{i \in \{1,\dots,m\}}L_{\mu_i}(\mathbb{X}), \quad
\omega_\sigma(\xi) = \max_{i \in \{1,\dots,m\}}\omega_{\sigma_i}(\xi),
\end{align*}for $i \in \{1,\dots,m\}$. Furthermore, for any $\delta \in (0,1)$ define
\begin{align*}
    \beta(\xi) = & 2\log \left(\frac{m M(\xi,\mathbb{X})}{\delta}  \right), \quad
    \gamma(\xi) = \left(\frac{L_f(\mathbb{X})}{n} + L_\mu(\mathbb{X})\right)\xi + \sqrt{\beta(\xi)}\omega_\sigma(\xi)  ,
\end{align*} where $M(\xi,\mathbb{X})$ is the $\xi$-covering  number of $\mathbb{X}$.
Then, we have
\[
\pr \left\{ \norm{f(x) - \mu(x)}_\infty \leq e_f(x) = \sqrt{\beta(\xi)}\norm{\sigma(x)}_\infty + \gamma(\xi), \quad \forall x \in \mathbb{X}  \right\} \geq 1-\delta.
\]
\end{theorem}  
\begin{proof}
The proof follows the arguments as in~\citet[Thm.~3.1]{lederer2019uniform} and is provided for completeness. We first establish the Lipschitz continuity of the mean function vector $\mu(x)$. For any $x,x' \in \mathbb{X}$, using the definition of $\mu_i$ in~\eqref{eqn:posterior_conditionals}, we obtain
\begin{equation}\label{eqn:proof:1}
    \left|\mu_i(x) - \mu_i(x')  \right| \leq \norm{\mathbf{K}_i^\star(x) - \mathbf{K}_i^\star(x')} \norm{\left( \mathbf{K}_i + \sigma_n^2 \mathbb{I}_N \right)^{-1}\mathbf{Y}_i}, \quad i \in \{1,\dots,m\}.
\end{equation} Using the Lipschitz continuity of the individual kernel functions in Assumption~\ref{assmp:kernel}, we get
\[
   \norm{\mathbf{K}_i^\star(x) - \mathbf{K}_i^\star(x')} \leq \sqrt{N}L_{K,i}(\mathbb{X})\norm{x - x'}, \quad \forall x,x' \in \mathbb{X}, \quad i \in \{1,\dots,m\}.
\] Thus, substituting in~\eqref{eqn:proof:1} produces
\[
   \left|\mu_i(x) - \mu_i(x')  \right| \leq L_{\mu_i}(\mathbb{X}) \norm{x - x'}, \quad \forall x,x' \in \mathbb{X}, \quad i \in \{1,\dots,m\},
\] which in turn implies
\begin{equation}\label{eqn:proof:2}
    \norm{\mu(x) - \mu(x')}_\infty \leq L_\mu(\mathbb{X})\norm{x - x'}, \quad \forall x,x' \in \mathbb{X}.
\end{equation}

We now establish the modulus of continuity of $\sigma(x)$. Using the non-negativity of $\sigma_i(x)$, we get
\begin{equation}\label{eqn:proof:3}
  \left|\sigma_i^2(x) - \sigma_i^2(x')  \right| \geq    \left|\sigma_i(x) - \sigma_i(x')  \right|^2,  \quad \forall x,x' \in \mathbb{X}, \quad i \in \{1,\dots,m\}.
\end{equation} Using the definition of $\sigma_i^2$ in~\eqref{eqn:posterior_conditionals}, we obtain
\begin{align}
    \left|\sigma_i^2(x) - \sigma_i^2(x')  \right| \leq & \left| \mathbf{K}_i^{\star \star}(x) - \mathbf{K}_i^{\star \star}(x')  \right| \notag \\
    &\label{eqn:proof:4}+ \norm{\mathbf{K}_i^\star(x) - \mathbf{K}_i^\star(x')}
     \norm{\left( \mathbf{K}_i + \sigma_n^2 \mathbb{I}_N \right)^{-1}}
     \norm{\mathbf{K}_i^\star(x) + \mathbf{K}_i^\star(x')},
\end{align} for all $x,x' \in \mathbb{X}$, and $i \in \{1,\dots,m\}$. The terms on the right hand side of the above expression can be bounded as
\begin{subequations}\label{eqn:proof:mod_cont_bounds}
\begin{align}
    \left| \mathbf{K}_i^{\star \star}(x) - \mathbf{K}_i^{\star \star}(x')  \right| \leq & 2 L_{K,i}(\mathbb{X})\norm{x - x'},\\
     \norm{\mathbf{K}_i^\star(x) - \mathbf{K}_i^\star(x')} \leq & \sqrt{N}L_{K,i}(\mathbb{X})\norm{x - x'}, \\
      \norm{\mathbf{K}_i^\star(x) + \mathbf{K}_i^\star(x')} \leq & 2 \sqrt{N}\max_{x,x' \in \mathbb{X}}K_{f,i}(x,x'),
\end{align} for all $x,x' \in \mathbb{X}$, and $i \in \{1,\dots,m\}$.
\end{subequations} Substituting~\eqref{eqn:proof:mod_cont_bounds} into~\eqref{eqn:proof:4} produces
\[
  \left|\sigma_i(x) - \sigma_i(x')  \right| \leq \sqrt{ \left|\sigma_i^2(x) - \sigma_i^2(x')  \right|}   \leq \omega_{\sigma_i}\left(\norm{x - x'} \right),
\] for all $x,x' \in \mathbb{X}$, and $i \in \{1,\dots,m\}$, where we have additionally used the inequality in~\eqref{eqn:proof:3}. Therefore, we conclude
\begin{equation}\label{eqn:proof:8}
    \norm{\sigma(x) - \sigma(x')}_\infty \leq \omega_\sigma\left(\norm{x - x'}\right), \quad \forall x,x' \in \mathbb{X}.
\end{equation}

We now establish prediction error bounds on sets of finite cardinality. Let $\mathbb{X}_\xi$ denote a countable discretization of the compact set $\mathbb{X}$ such that
\begin{equation}\label{eqn:proof:9}
|\mathbb{X}_\xi| < \infty \quad \text{and} \quad \max_{x \in \mathbb{X}} \min_{x' \in \mathbb{X}_\xi}\norm{x - x'} \leq \xi.    
\end{equation} Using the posterior distribution of $f_i$ in~\eqref{eqn:posterior_conditionals}, we have that
\[
 \frac{1}{\sigma_i(x)}\left(f_i(x) - \mu_i(x) \right) \sim \mathcal{N}(0,1), \quad \forall x \in \mathbb{X}_\xi, \quad i \in \{1,\dots,m\}.
\] Then, from~\citet[Lemma~5.1]{srinivas2012information}, we have that for any $x \in \mathbb{X}_\xi$ and $i \in \{1,\dots,m\}$, the following holds
\[
 \pr \left\{ |f_i(x) - \mu_i(x)| > \sqrt{\beta(\xi)} \sigma_i(x)  \right\} \leq e^{-\beta(\xi)/2}.
\] Applying the union bound over the set $\mathbb{X}_\xi \times \{1,\dots,m\}$, we conclude that
\[
  |f_i(x) - \mu_i(x)| \leq \sqrt{\beta(\xi)} \sigma_i(x), \quad \forall x \in \mathbb{X}_\xi, \quad i \in \{1,\dots,m\}
\] holds with the probability of at least $1 - m|\mathbb{X}_\xi|e^{-\beta(\xi)/2}$. Using the definition of $\beta(\xi)$, we get that
\[
  |f_i(x) - \mu_i(x)| \leq \sqrt{\beta(\xi)} \sigma_i(x), \quad \forall x \in \mathbb{X}_\xi, \quad i \in \{1,\dots,m\},
\] holds with the probability of at least $1 - \delta$. Therefore, we have
\begin{equation}\label{eqn:proof:10}
    \pr \left\{ \norm{f(x) - \mu(x)}_\infty \leq \sqrt{\beta(\xi)} \norm{\sigma(x)}_\infty, \quad \forall x \in \mathbb{X}_\xi     \right\} \geq 1 - \delta.
\end{equation}

Using the Lipschitz continuity of $f(x)$ and $\mu(x)$ in Assumption~\ref{assmp:Lip_bounds} and Equation~\eqref{eqn:proof:2}, respectively, and the modulus of continuity of $\sigma(x)$ in~\eqref{eqn:proof:8}, we obtain that for all $x \in \mathbb{X}$ and $x' \in \mathbb{X}_\xi$
\begin{subequations}
\begin{align}
    \norm{f(x) - f(x')}_\infty \leq &\label{eqn:proof:11a} \frac{L_f(\mathbb{X})}{n} \norm{x - x'},\\
    \norm{\mu(x) - \mu(x')}_\infty \leq &\label{eqn:proof:11b} L_\mu(\mathbb{X}) \norm{x - x'},\\
    \norm{\sigma(x) - \sigma(x')}_\infty \leq &\label{eqn:proof:11c} \omega_\sigma(\norm{x - x'}).
\end{align}
\end{subequations} Next,  we have
\begin{align*}
    \norm{f(x) - \mu(x)}_\infty \leq & \norm{f(x) - f(x')}_\infty + \norm{\mu(x) - \mu(x')}_\infty + \norm{f(x') - \mu(x')}_\infty
\end{align*} for all $x \in \mathbb{X}$ and $x' \in \mathbb{X}_\xi$. Using~\eqref{eqn:proof:10},~\eqref{eqn:proof:11a}-\eqref{eqn:proof:11b}, we get
that
\begin{align}
    \norm{f(x) - \mu(x)}_\infty \leq &\label{eqn:proof:12} \left(\frac{L_f(\mathbb{X})}{n} +  L_\mu(\mathbb{X}) \right)\norm{x - x'} + \sqrt{\beta(\xi)}\norm{\sigma(x')}_\infty, 
\end{align} for all $x \in \mathbb{X}$ and $x' \in \mathbb{X}_\xi$ holds with the probability of at least $1- \delta$. Note that
\[
\norm{\sigma(x')}_\infty \leq \norm{\sigma(x') - \sigma(x)}_\infty + \norm{\sigma(x)}_\infty
\] for all $x \in \mathbb{X}$ and $x' \in \mathbb{X}_\xi$. The use of~\eqref{eqn:proof:8} and~\eqref{eqn:proof:11c} leads to the conclusion that
\[
   \norm{\sigma(x')}_\infty \leq \omega_\sigma(\norm{x - x'}) + \norm{\sigma(x)}_\infty,
\] for all $x \in \mathbb{X}$ and $x' \in \mathbb{X}_\xi$. Substituting into~\eqref{eqn:proof:12} leads to the conclusion that
\[
   \norm{f(x) - \mu(x)}_\infty \leq  \left(\frac{L_f(\mathbb{X})}{n} +  L_\mu(\mathbb{X}) \right)\norm{x - x'} + \sqrt{\beta(\xi)}\omega_\sigma(\norm{x - x'})      + \sqrt{\beta(\xi)}     \norm{\sigma(x)}_\infty 
\] for all $x \in \mathbb{X}$ and $x' \in \mathbb{X}_\xi$ holds with the probability of  at least $1- \delta$. Finally, using~\eqref{eqn:proof:9}  completes the proof.

\end{proof}

\subsection{Overview of $\mathcal{L}_1$ Adaptive Control}\label{subsec:L1_overview}


In this subsection, we briefly review the existing standard \lonew control architecture for the uncertain system \eqref{eqn:system_dynamics} {\it without} incorporation of learned dynamics. Consequently, in Section~\ref{sec:L1-GP} we will show how the GPR learned dynamics can be incorporated within the $\mathcal{L}_1$ architecture. The reader is directed to \citet{L1_book}, especially its Section 3.3, for further details on the following material.  An \lonew controller mainly consists of three components: a state predictor, an adaptation law,  and a control law. The state predictor is used to generate an estimate of the tracking error, which is subsequently used in the adaptation laws to update the uncertainty estimates. We consider the piecewise-constant adaptation law that is inherently connected with the CPU sampling rate. The control law  cancels the estimated uncertainty within the bandwidth of the low-pass filter.
For the uncertain system \eqref{eqn:system_dynamics}, these components are detailed as follows. 
The \textbf{state predictor} is given as
\begin{equation}\label{eqn:vanilla_predictor}
\dot{\hat{x}}(t) = A_m \hat{x}(t) + B_m(u(t) + \hat{\sigma}(t)), \ \hat{x}(0) = \hat{x}_0, \quad \text{and} \quad \hat{y}(t) =  C_m \hat{x}(t),
\end{equation}
where $\hat{x}(t)\in \mathbb{R}^n$ is the predictor state and $\hat{x}_0$ is its initial value (that may be different from $x_0$ in \eqref{eqn:system_dynamics}),
$\hat{\sigma}(t)\in \mathbb{R}^m$
is the adaptive estimate.
The \textbf{adaptive estimate} is  updated according to
\begin{equation}\label{eqn:vanilla_adaptation}
\begin{split}
\hat{\sigma}(t) 
& = \hat{\sigma}(iT_s), \quad  \hat{\sigma}(iT_s) 
 = -
B_m^{+}\Phi^{-1}(T_s)e^{A_mT_s}\tilde{x}(iT_s),
\end{split}
\end{equation}
where  $ t\in [iT_s, (i+1)T_s] $  with $T_s$ being the  sampling time and $i\in \cal{Z}_{+} $. In addition, $B_m^{+} = (B_m^TB_m)^{-1}B_m^T$ is the pseudo-inverse of $B_m$, $\Phi(T_s) \triangleq A_m^{-1}(e^{A_mT_s}-\mI_n)$, and 
$\Tilde{x}(t) \triangleq \hat{x}(t) - x(t)$ is the prediction error. 
The \textbf{control law} is given as
\begin{equation}\label{eqn:vanilla_control_law}
u(s) = C(s)(\hat{\sigma}(s) - k_g r(s)),    
\end{equation} where $\hat{\sigma}(s)$ is the Laplace transform of $\hat{\sigma}(t)$, $r(t)$ is the reference signal and $k_g\triangleq -(C_mA_m^{-1}B_m)^{-1}$ is a feedforward gain to ensure that the desired transfer function matrix $M(s) = C_m(s\mI_n-A_m)^{-1}B_m$ has DC gain equal to an identity matrix, and $C(s)$ is a lowpass filter with $C(0)= \mI_m$, subject to the following $\mathcal L_1$-norm condition:
\begin{equation}\label{eqn:L1norm}
 \norm{H(s)(\mI-C(s))}_{\lone} < \frac{\rho_r - \norm{H(s)C(s)k_g}_\lone \norm{r}_\linf-\rho_\textup{in}}{L_{f}(\mX_{\rho_r})\rho_r + B_0},
\end{equation}
where $H(s)\triangleq  (s\mI_n-A_m)^{-1}B_m$, $\rho_\textup{in} \triangleq \norm{s(s\mI-A_m)^{-1}}_\lone \rho_0$  with $\rho_0$ being a known bound for the initial state $x_0$ (i.e. $\norm{x_0}_\infty\leq \rho_0$), $B_0$ and $L_f(\cdot)$ are defined in Assumption~\ref{assmp:Lip_bounds}, $\rho_r$ is a positive constant that defines the semiglobal domain of attraction. The reference model and filter can be designed via optimization~\citep{jafarnejadsani2017optimized}, however the best way to perform this optimization is still an open problem. Heuristic design choices can be found in~\citet[Section 2.6]{L1_book}. 
When there is no initialization error, i.e. $\hat{x}_0 = x_0$, following \citet{L1_book}, if $T_s\rightarrow 0$, then the state and control signals of the closed-loop $\mathcal L_1$ system -- both in transient and steady-state -- can be made arbitrarily close to the corresponding signals of the following  non-adaptive auxiliary reference system 
\begin{subequations}\label{eqn:reference_system}
\begin{align}
    \dot{x}_{\text{ref}}(t) = & A_m x_{\text{ref}}(t) + B_m ( u_{\text{ref}}(t) + f(x_{\text{ref}}(t))), \quad x_{\text{ref}}(0) = x_0,\\
    u_{\text{ref}}(s) = &C (s)(k_gr(s) - \eta_{\text{ref}}(s)),\quad y_{\text{ref}}(t)= C_m x_{\text{ref}}(t),
\end{align}
\end{subequations} where $\eta_{\text{ref}}(s)$ is the Laplace transform of $\eta_{\text{ref}}(t) \triangleq f(x_{\text{ref}}(t))$. In the presence of non-zero initialization error, the  performance bounds between the adaptive system and the reference system will  contain additive exponentially decaying terms that depend on the initialization error.
The reference system defines the \textit{ideal achievable performance}, where the uncertainty is perfectly known and cancelled within the bandwidth of the filter $C(s)$. Its stability hinges upon the same condition in \eqref{eqn:L1norm}, while the bandwidth of the filter $C(s)$ defines the tradeoff between performance and robustness.

\section{The $\mathcal{L}_1$-$\mathcal{GP}$ Architecture} \label{sec:L1-GP}

The architecture of the $\mathcal{L}_1$-$\mathcal{GP}$ controller contains two primary components: i) the Bayesian learner that uses a GPR algorithm to produce estimates of the uncertainty $f$, and ii) the $\mathcal{L}_1$ adaptive  controller which incorporates the estimates and generates the control input $u(t)$. 

\noindent \textbf{Bayesian learner:}\label{subsec:Bayesian_lerner} The task of the Bayesian learner is to use the collected data to produce the estimates of the uncertainty $f$ in the form of the mean function $\mu$ of the posterior distribution. Furthermore, it also outputs the high-probability prediction error bounds presented in Theorem~\ref{theorem:uniform_bounds}. The output of the Bayesian learner is given by
\begin{equation}\label{eqn:model_updates}
\mathcal{M}(x(t),t) = \{\hat{f}(x(t),t),\hat{e}_f(x(t),t)\},
\end{equation} where the piecewise static in time $\hat{f}$ and $\hat{e}$ are defined as $\hat{f}(x(t),t) = \mu_k(x(t))$ and $\hat{e}_f(x(t),t) = e_{f,k}(x(t))$, for all $t \in [t_k,t_{k+1})$, $t_k \in \mathcal{T}$. 
Here, $\mathcal{T}$ is the set of discrete time-instances at which the Bayesian learner updates the model parameters. Thus, over the time interval $[t_k,t_{k+1})$, $\mu_k(x(t)) = \begin{bmatrix}\mu_{k,1}(x(t)) & \cdots & \mu_{k,m}(x(t))   \end{bmatrix}$, where $\mu_{k,i}(\cdot)$ are the mean functions obtained after the $k^{\text{th}}$-model update computed via the posterior distributions in~\eqref{eqn:posterior_conditionals}. Similarly, $e_{f,k}(x(t))$ is the uniform error bound computed via Theorem~\ref{theorem:uniform_bounds} after the $k^{\text{th}}$ model update. The Bayesian learner updates the model once $N \in \mathbb{N}$ new data points have been collected; thus $N$ is a design parameter. The Bayesian learner is initialized to $\mu_{0}(x(t)) = 0_m$, which is the prior mean, and $e_{f_0}(x(t)) = e_f(x(t))$ is obtained based solely on the GP priors on $f$. 


\noindent \textbf{Incorporating Learning into $\mathcal{L}_1$ Control:} Next, we present the $\mathcal{L}_1$-$\mathcal{GP}$ controller that incorporates the model updates produced by the Bayesian learner into the $\mathcal{L}_1$ controller. Same as the $\mathcal{L}_1$ controller, the $\mathcal{L}_1$-$\mathcal{GP}$ controller consists of the state-predictor, adaptation law, and the control law. The $\mathcal{L}_1$-$\mathcal{GP}$ state-predictor is given by
\begin{align}
 \dot{\hat{x}}(t) =& A_m \hat{x}(t) + B_m \left(f_L(t) + \hat{\sigma}(t)  + u(t)  \right),~~ \hat{x}(0) = \hat{x}_0 ~~ \text{and}  \quad\hat{y}(t) =  C_m \hat{x}(t),\label{eqn:L1GP_predictor}
\end{align} 
where $\hat{\sigma}(t)$ is the adaptive estimate of uncertainties, 
$f_L(t)$ is the solution of the following equation 
\begin{equation}\label{eqn:learning_filter}
\dot{f}_{L}(t) = -\omega(t)\left(f_{L}(t) - \hat{f}(x(t),t)  \right), \quad f_{L}(0) = 0,    
\end{equation}  with $\hat{f}(x(t),t)$ being defined in~\eqref{eqn:model_updates}, and
\begin{equation}\label{eqn:adaptive_bandwidth}
\omega(s) = L(s)\hat{\omega}(s), \quad \hat{\omega}(t) = \min \left\{\omega_0/\hat{e}_f(x(t),t),\omega_c\right\}.
\end{equation} Here, $\omega_0$ is an an arbitrarily small \textit{apriori} chosen positive scalar, and $\omega_c$ 
is the bandwidth of $C(s)$  verifying the $\mathcal{L}_1$-norm condition in \eqref{eqn:L1norm},  
$\hat{e}_f(x(t),t)$ is the output of the Bayesian learner defined in~\eqref{eqn:model_updates}, and $L(s)$ is a low-pass filter. 
The update of the adaptive estimate $\hat{\sigma}$ is governed by the piecewise-constant adaptation law with sampling time $T_s$ as defined in~\eqref{eqn:vanilla_adaptation}. Finally, the $\mathcal{L}_1$-$\mathcal{GP}$ control law is given by 
\begin{equation}\label{eqn:L1GP_control_law}
u(s) = -f_{L}(s) - C(s)(\hat{\sigma}(s) - k_g r(s)).   
\end{equation} 
 Note that $\hat{e}_f(x(t),t)$, defined in~\eqref{eqn:model_updates}, starts at $e_{f_0}(x(t))$ when no model updates have been performed, and ideally approaches zero after sufficiently large number of 
model updates have been performed as the size of the data set increases. 
Therefore, by the law presented in~\eqref{eqn:adaptive_bandwidth}, $\omega(t)$  
in~\eqref{eqn:learning_filter} increases from an arbitrarily small value $\omega_0/e_{f_0}(x(t))$ to $\omega_c$, the bandwidth of the filter $C(s)$. Moreover, the change in $\omega(t)$ is smooth because of the low-pass filter $L(s)$. In this way the filter~\eqref{eqn:learning_filter} allows the incorporation of the learned uncertainties smoothly into the system. 
In addition, as $\hat{f} \rightarrow f$\footnote{The expression $\hat{f} \rightarrow f$ implies that the high-probability bounds on $\|f(x) - \mu(x)\|_\infty$ go to zero. The conditions under which this convergence takes place can be found in~\citet{lederer2019uniform}.} 
, it is to be expected that $\tilde{x}(t)$ and $\hat{\sigma}(t)$ go to zero. Thus, the $\mathcal{L}_1$-$\mathcal{GP}$ closed-loop system defined by~\eqref{eqn:system_dynamics},~\eqref{eqn:L1GP_predictor}-\eqref{eqn:L1GP_control_law} converges to the $\mathcal{L}_1$ reference system in~\eqref{eqn:reference_system}. 
The adaptive estimate $\hat{\sigma}$ is driven by the prediction error $\tilde{x} \triangleq \hat{x} - x$, whose evolution
is governed by
\begin{equation}\label{eqn:L1GP:pred-error}
    \dot{\tilde{x}}(t) = A_m \tilde{x}(t) + B_m \left(f_{L}(t) - f(x(t)) + \hat{\sigma}(t)  \right), \quad \tilde{x}(0) = \hat{x}_0 - x_0.
\end{equation}
 The learned dynamics are used to cancel the model uncertainty via $f_L(t)$ in~\eqref{eqn:learning_filter}. From the prediction error dynamics \eqref{eqn:L1GP:pred-error}, it is evident that the $-C(s)\hat{\sigma}(s)$ component of the control law~\eqref{eqn:L1GP_control_law} compensates for the remaining uncertainty, $f(x(t))-f_L(t)$, within the bandwidth of the filter $C(s)$. 
 \begin{remark}
 Proof of the stability of the $\mathcal{L}_1$-$\mathcal{GP}$ closed-loop system can be established by following the ideas in \citet{cooper2014learning,snyder2019phd}.  \end{remark}
\section{Simulation Results}\label{sec:sims}
\begin{figure}[t]
\vspace{-7mm}
\begin{center}
\begin{tabular}{cc}
\subfigure[State evolution ]{\includegraphics[width =0.45\textwidth]{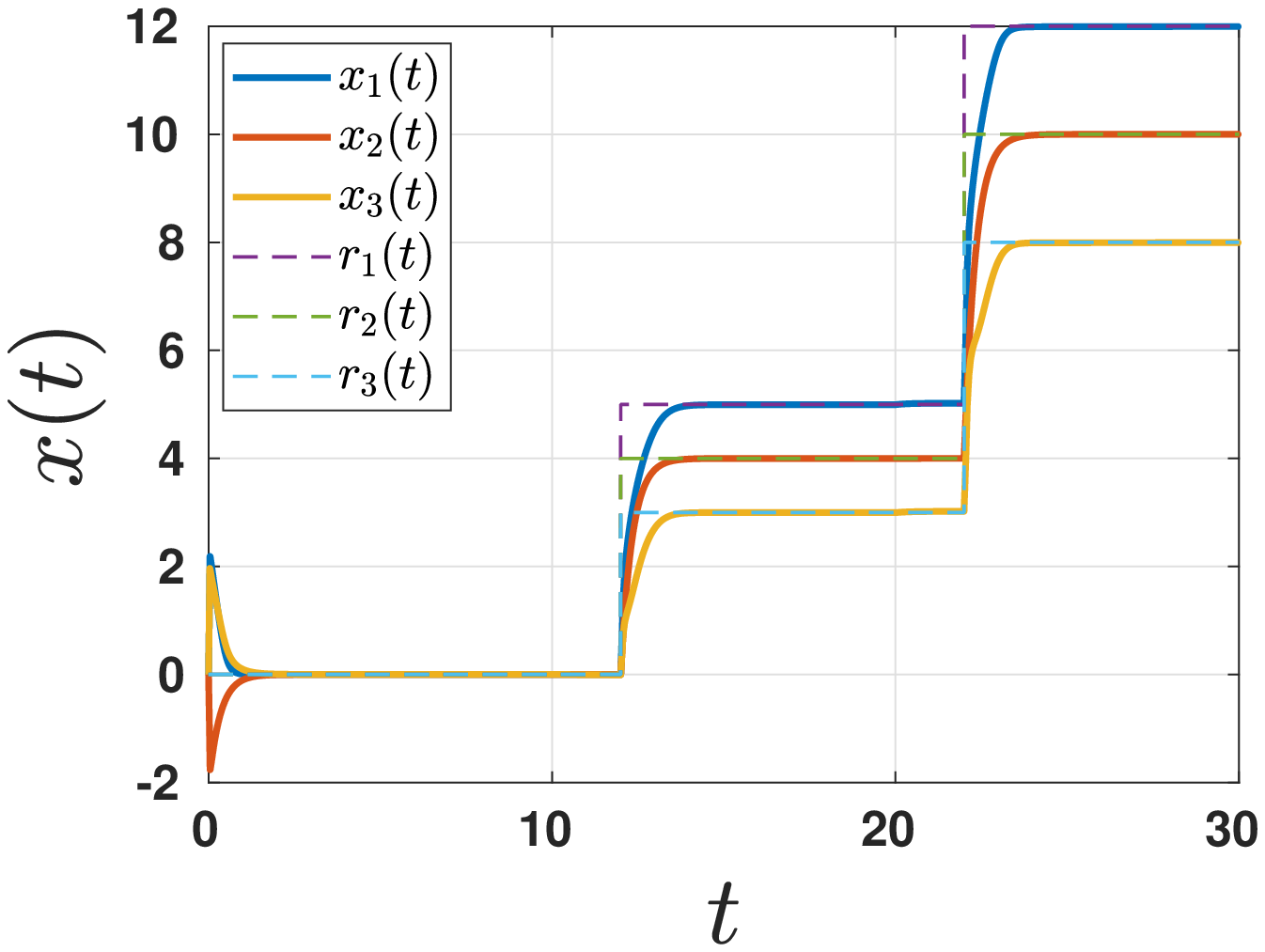}} & 
\subfigure[Control int evolution]{\includegraphics[width = 0.45\textwidth]{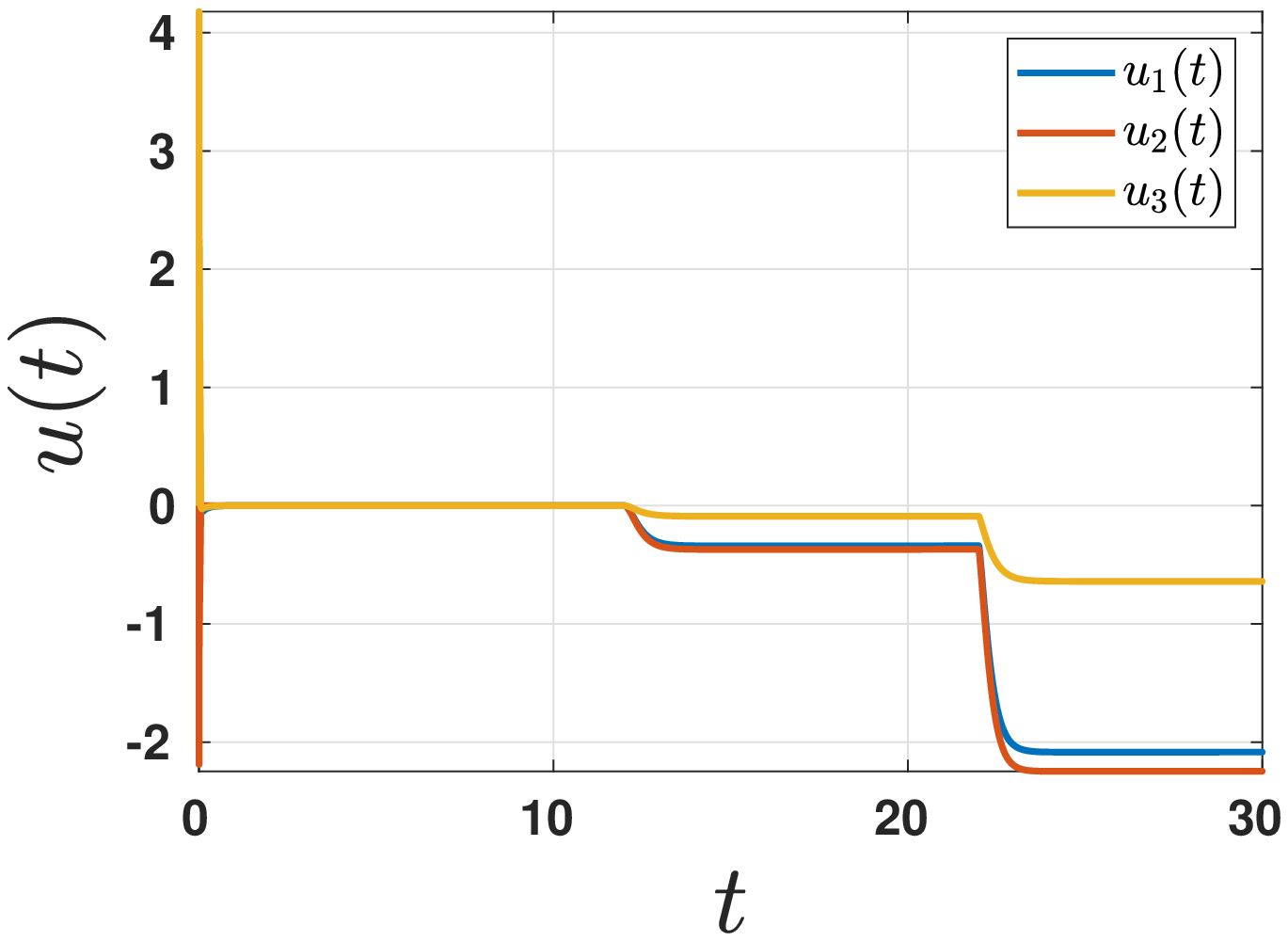}}
\end{tabular}
\end{center}
\vspace{-7mm}
\caption{State and control input evolution for $\mathcal{L}_1$-$\mathcal{GP}$ closed-loop system for step reference inputs.}
\label{fig:state_and_control}
\vspace{-3mm}
\end{figure}

We now present the results of numerical experimentation. We consider the dynamics of body-frame angular rates $x(t) \in \mathbb{R}^3$ of a multirotor craft given by
\begin{subequations}\label{eqn:rate_dynamics}
\begin{align}
   \dot{x}(t) =& -J^{-1}\left(x(t) \times J x(t)  \right) + J^{-1}f(x(t)) + J^{-1}u_{total}(t), \quad x(0) = x_0 = 0_3,\\ y(t) = & x(t),
\end{align}
\end{subequations} where $J = \text{diag}\{0.011,0.011,0.021\}$ is the known moment-of-inertia matrix, $f(x(t))$ is the model uncertainty, and $u_{total}(t) \in \mathbb{R}^3$ is the control input, which, for a multirotor craft presents the body-frame moments. The control input is decomposed as $u_{total}(t) = u_{bl}(t) + u(t)$, where $u_{bl}(t)$ is the baseline input and $u(t)$ is the $\mathcal{L}_1$-$\mathcal{GP}$ input. The role of the baseline input is to inject desired dynamics, i.e., $u_{bl}(t) = JA_m x(t) + \left(x(t) \times J x(t)  \right)$, where $A_m = -3 \mathbb{I}_3$. With baseline input injected into~\eqref{eqn:rate_dynamics}, the partially closed-loop system can be written in the form of~\eqref{eqn:system_dynamics} with $B_m = J^{-1}$ and $C_m = \mathbb{I}_3$. Next, we consider the following model uncertainty
\begin{equation}\label{eqn:model_uncertainty}
f(x(t)) =  \begin{bmatrix}
0.01\left( x_1^2(t) + x_3^2(t) \right) & 0.01\left(x_3(t)x_2(t) + x_1^2(t)  \right) & 0.01\left( x_3^2(t)  \right)
\end{bmatrix}^\top.    
\end{equation}
For the $\mathcal{L}_1$-$\mathcal{GP}$ control input, we set $C(s) = \omega_c/(\omega_c + s) \mI_3$, $\omega_c = 80~rad/s$, $L(s) = 0.01/(0.01 + s)$, and $\omega_0 = 1$. The predictor~\eqref{eqn:L1GP_predictor} is initialized with $\hat{x}_0 = \begin{bmatrix} 0.5 & 0.5 & 0.5  \end{bmatrix}^\top$, which is distinct from the system's initial conditions in~\eqref{eqn:rate_dynamics}. For the GPR, we choose the Squared-Exponential (SE) kernels as $K_{f,i}(x,x') = \sigma_f^2 \text{exp}\left(-(x-x')^\top (x-x')/2l^2  \right)$, where the unoptimized hyper-parameters are chosen to be $\sigma_f = l = 1$. Furthermore, we upper bound the covering number $\beta(\xi)$ (Thm.~\ref{theorem:uniform_bounds}) as in~\citet{lederer2019uniform} using $\xi = 0.001$ and conservatively chosen $\mathbb{X} = \{x \in \mathbb{R}^3~:~\|x\|_\infty \leq 15\}$. For the purposes of simulation, we ignore the $\gamma(\xi)$ term (Thm.~\ref{theorem:uniform_bounds}) as they can be made arbitrarily small. Finally, we choose $\delta = 0.01$, the feedforward gain $k_g = -\left(C_m A_m^{-1}B_m  \right)^{-1}$ and the sampling time for the update of the adaptive estimate $\hat{\sigma}(t)$ as $T_s = 0.001$. 
The Bayesian learner collects data at the rate $1~Hz$ and updates the model after $N=10$ new data-points have been collected; thus the model is updated at $0.1~Hz$. Figure~\ref{fig:state_and_control} illustrates the state evolution and the $\mathcal{L}_1$-$\mathcal{GP}$ input $u$ in response to a step reference command. The figure shows the scaled response of the system without retuning, a property that $\mathcal{L}_1$-$\mathcal{GP}$ shares with $\mathcal{L}_1$ control. Moreover, $\mathcal{L}_1$-$\mathcal{GP}$ preserves the performance bounds which are guaranteed for $\mathcal{L}_1$ control.  
\begin{wrapfigure}{r}{0.4\textwidth}
\vspace{-1mm}
    \includegraphics[width=0.43\textwidth]{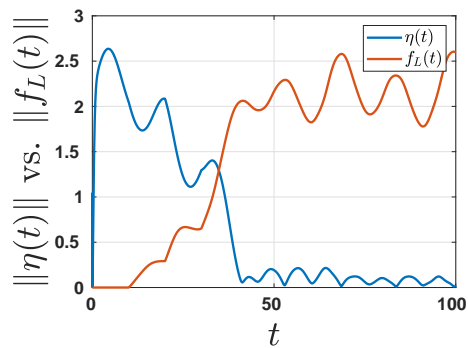}
    \centering
       \vspace{-7mm} 
    \caption{Evolution of $\|f_L(t)\|$ and $\|\eta(t)\|$ for sinusoidal reference commands, where $\eta(s) = C(s)\hat{\sigma}(s)$.}
    \label{fig:learning_v_adaptive}
 \vspace{-4mm} 
\end{wrapfigure}
Next we show the effect of learning within the $\mathcal{L}_1$-$\mathcal{GP}$ input $u(t)$. Recall that $u(t)$ in~\eqref{eqn:L1GP_control_law} is comprised of two major components, the learning based input $f_L(t)$ and the adaptive input $\eta(t)$, where $\eta(s) = C(s) \hat{\sigma}(s)$. The evolution of these individual components for a sinusoidal reference is illustrated in Figure~\ref{fig:learning_v_adaptive}. 
Note that the dominant component of the control input $u(t)$ transitions from adaptive input $\eta(t)$ to the learning based input $f_L(t)$ as the learning improves.

\begin{figure}[htp]
\vspace{-7mm}
\begin{center}
\begin{tabular}{cc}
\subfigure[Evolution of $\|f_L(t)\|$ and $\|\eta(t)\|$.]{\includegraphics[width = 0.45\textwidth]{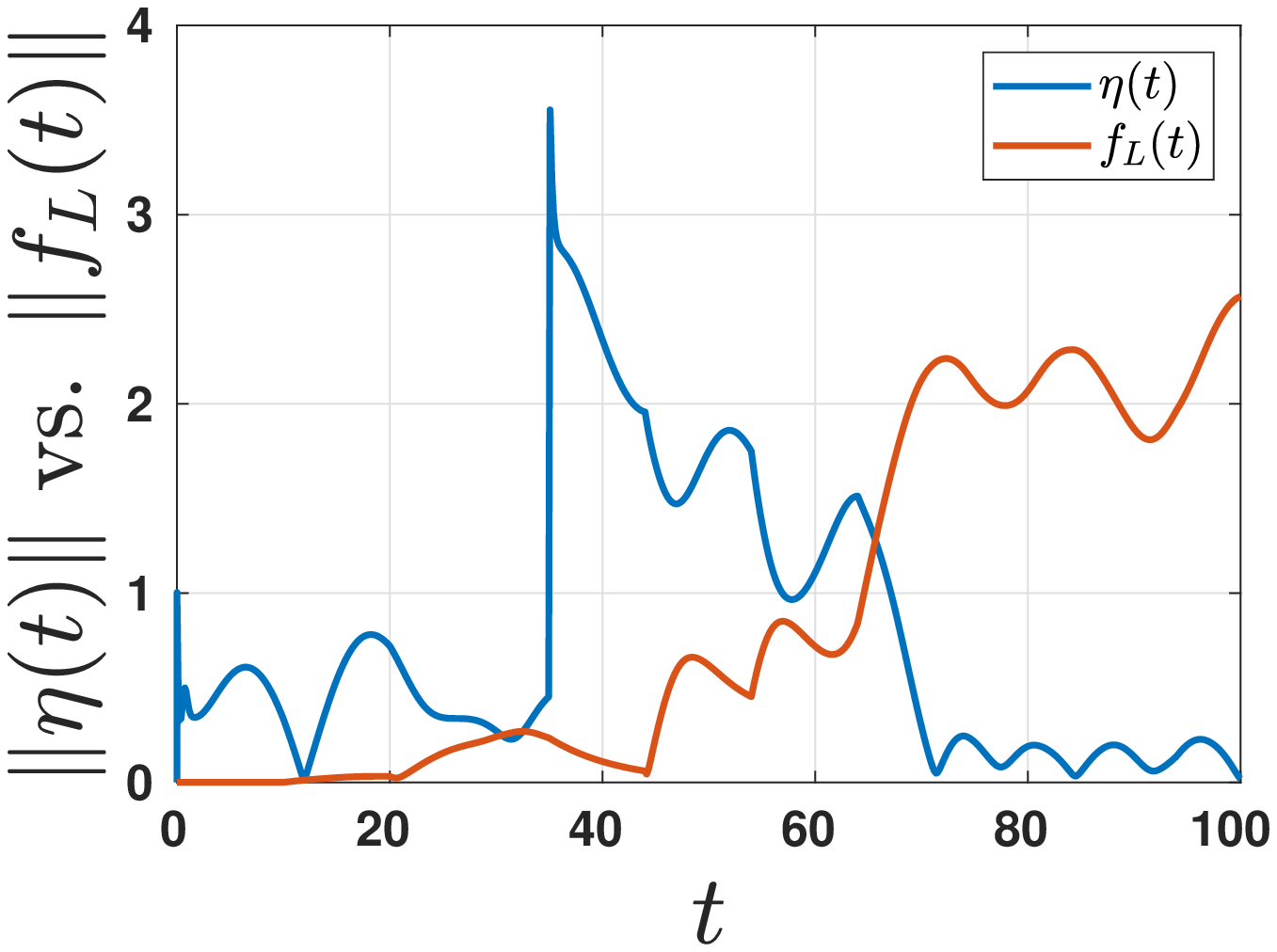}} & 
\subfigure[State evolution. Inset shows the smooth response of the system state across the uncertainty switch.]{\includegraphics[width = 0.45\textwidth]{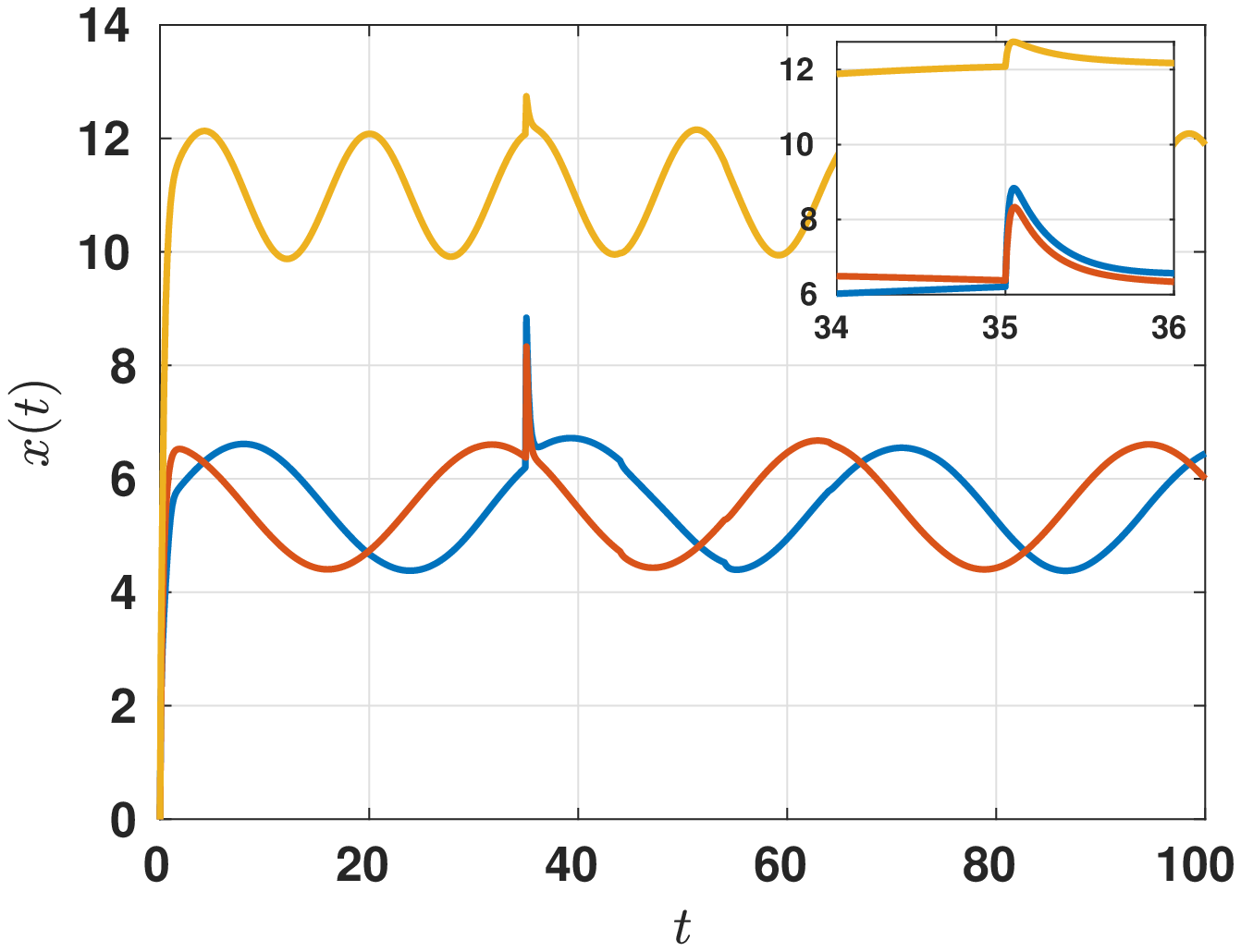}}
\end{tabular}
\end{center}
\vspace{-7mm}
\caption{Learning and adaptive components of the $\mathcal{L}_1$-$\mathcal{GP}$ input $u(t)$ and system state evolution with model uncertainty switch at $t = 35s$.}
\label{fig:disturbance}
\vspace{-3mm}
\end{figure}

We now demonstrate the safe-learning enabled by the $\mathcal{L}_1$-$\mathcal{GP}$ controller under sudden change of uncertainties. As illustrated in Figure~\ref{fig:learning_v_adaptive}, as the learning improves, the learning based component $f_L(t)$ becomes the major contributor to $u(t)$. However, the adaptive component, $\eta(t)$, always remains active in the background ready to intervene when new uncertainties enter the dynamics. This is crucial for stability and performance guarantees 
as the learning runs on a long time scale, whereas the fast adaptation due to $\hat{\sigma}(t)$ immediately intervenes to compensate for the new uncertainties. To demonstrate this, the $\mathcal{L}_1$-$\mathcal{GP}$ controller is tasked with tracking a sinusoidal reference command. At $t = 35~s$, we switch the model uncertainty from $f(x(t))$ in~\eqref{eqn:model_uncertainty} to
$
f(x(t)) = \left[  0.5 \sin (x_1(t))\ 0.01 \cos (x_3(t)) \ 0.5 \left(\sin (x_1(t)) + \cos(x_2(t))  \right)   \right]^\top.
$ The results are illustrated in Figure~\ref{fig:disturbance}. At $t = 35~s$, when the uncertainty $f(x(t))$ switches, the adaptive element $\eta(t)$ immediately intervenes to compensate for the new uncertainty. Furthermore, at this point, the previously learned input $f_L(t)$ is incapable of cancelling the new $f(x(t))$. Therefore, $\eta(t)$ considers $f_L(t)$ as a disturbance to be rejected. However, since $f_L(t)$ enters the system via the low-pass filter~\eqref{eqn:learning_filter}, it always remains within the bandwidth of $C(s)$, and thus can be compensated by the  adaptive element $\eta(t)$. 
Finally, the state evolution illustrates the maintenance of stability of the system. 

We would also like to remark that both the $\mathcal{L}_1$-$\mathcal{GP}$  and the $\mathcal{L}_1$ control maintain the same time-delay margins. The time-delay margins for both  control schemes were computed numerically to be $\approx20~ms$. This is not surprising since the time-delay margins are dominated by the adaptive elements including the low-pass filter $C(s)$ and sampling time $T_s$, which are the same for the $\mathcal{L}_1$-$\mathcal{GP}$ and the $\mathcal{L}_1$ controllers.

\section{Conclusion}\label{sec:conclsn}
We presented the $\mathcal{L}_1$-$\mathcal{GP}$ architecture, which incorporates Bayesian learning via Gaussian Process Regression (GPR) into the $\mathcal{L}_1$ adaptive control framework. Within the framework, GPR allows for sample-efficient learning of the model uncertainties, while the $\mathcal{L}_1$ controller provides stability, robustness and performance guarantees throughout the learning phase. 
We demonstrated the efficacy of the proposed architecture through numerical simulations. 
The $\mathcal{L}_1$-$\mathcal{GP}$ architecture is the initial phase of the research  and will next proceed by using learning to improve the performance over a larger envelope of operation, while maintaining given robustness specifications.
Eventually, the presented work will be extended to safe and robust planning and control of uncertain systems.
The $\mathcal{L}_1$-$\mathcal{GP}$ architecture will be extended to consider spatio-temporal learning for realistic scenarios as most real systems are subject to time-varying disturbances.
Further extensions of the $\mathcal{L}_1$-$\mathcal{GP}$ architecture to the case of output-feedback and stochastic systems will also be investigated.

This work is financially supported by the National Aeronautics and Space Administration (NASA) and National Science Foundation’s Cyber Physical Systems (CPS) award \#1932529.




\bibliography{thoughts,refs_pan,refs-naira}

\end{document}